WHITE PAPER 2016

# 5G FRAMEWORK CONCEPTS FOR THE NEXT GENERATION NETWORKS

Mobile communication technologies have been evolving for many years with each generation transforming the way we experience new services. As the Smartphone market has significantly expanded in recent years and expected to grow more in years to come, the network evolution must continue to keep up the pace with users' demand even beyond the common usage connectivity. The envisioned market space for the next generation technology is driven by requirements to enhance mobile broadband connectivity, reach a massive range of machine type communication (MTC), and target services with ultra-reliable and low latency (URLLC) communications. To deliver these requirements, 5G must be designed with scalability and diversity across many components from spectrum, core network, radio access and devices.

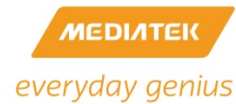

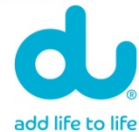

**Jointly Authored by: Ayman Elnashar[1] and Mohamed Elsaidny[2]**

[1] Head – Core and Cloud Planning (du)
[2] Sr. Regional Manager – Carriers Account at MediaTek

## CONTENTS

## 1. 5G EVOLUTION CONCEPTS: GETTING READY FOR 5G

The envisioned market space for 5G technology is targeting a design that has the capability to unify the system needed by various use cases. New use cases keep evolving according to the need for higher peak data rates, reduced end-to-end service latency, and increased network capacity in terms of user traffic and density. In addition, use cases suitable for the commercial 4G LTE network capabilities keep arising beyond today's usage and potentially even beyond what the network and devices are designed for. A wide range of new applications and use cases require advanced connectivity capabilities that can be both intense and diverse by its own requirements, including high capacity and data rate connectivity (3D video and virtual reality connectivity), real-time communications with low latencies (interactive video, automotive, critical type control, and tactile Internet), and massive Internet-of-Things connectivity (sensor networks, smart metering).

### Definition & Use Cases for 5G

At the moment, whenever any potential market space opens up, it require significant design change to the existing technologies, and in some cases, a total new radio access and core network. For example, when the M2M type applications for different types of industries started to materialize, a total new Internet-of-Things technology design was realized to be needed which did not help in accelerating the deployment of new services. A revamped radio access called Narrowband Internet of Things, NB-IOT, had to be designed by 3GPP to address the emerging adjacent industries.

All these factors require a new system designed to handle any new use case at any time without the need to re-dimension, re-design or even invest heavily in a total new network and technology components for each and single use case. 5G is promising on delivering a unified system that can be considered all-in-one; more than just developing new radio technology, it is for new possibilities and use cases. It aims at accelerating new business cases that keep promoting continuous costly and non-trivial changes into the existing ecosystem, from new cellular towers, to investing in new cloud computations and core networks, and devices that users can efficiently utilize for their everyday life.

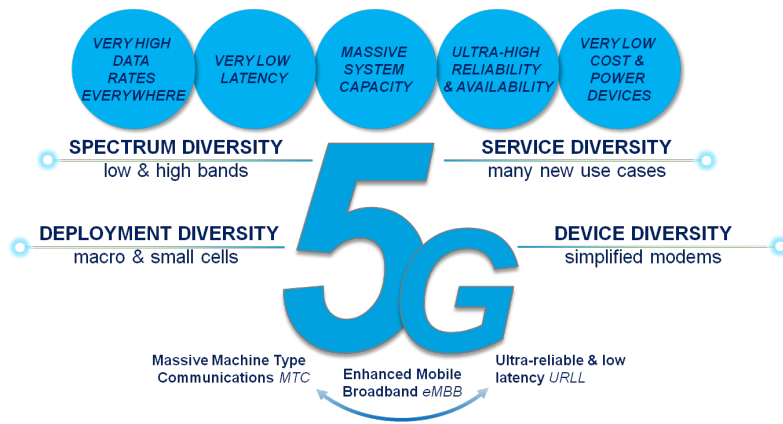

**FIGURE 1: 5G AND THE INDUSTRY DRIVING FORCE – ALL-IN-ONE SYSTEM**

In order to enable services for a wide range of users and industries with new requirements, the capabilities of 5G must extend beyond those of previous wireless access generations. These capabilities will include enhanced mobile broadband connectivity, massive system capacity for machine type communications, very high data rates everywhere, very low latency, ultra-high reliability and availability, and low device cost, as figure 1 shows.

## LTE Evolution to Enable 5G Use Cases

The major differences expected in 5G compared to legacy 4G generation will not only be at the level of combining an old and new radio access technologies; 5G will also enable new use cases and requirements of mobile communication beyond 4G systems. It will be an integration of existing cellular standards and technologies, including new disruptive technologies like millimeter-wave (mmWave) and spectrum sharing among other new concepts like network slicing. These concepts will facilitate the integration of vertical industries into the mobile ecosystem, whilst opening up new business models and revenue streams for operators.

Mobile broadband access and service availability with low latencies are key use cases driving the requirements for 5G, especially at the initial phase of deployment. The key enabler for any new design in 5G will be the spectrum and how to utilize concepts from previous technologies in order to accelerate the migration to 5G.

In order to meet the 5G requirements and bring this visionary system to reality, the future 5G network will be one that is built upon the small cell backbone either in standalone or non-standalone deployment. As spectrum suitable for mobile communication becomes more and more scarce, densification is the only way to meet the area traffic capacity demand. Even for millimeter wave band where spectrum is abundant, the channel's propagation characteristics will likely limit its range for mobile access to that of a small cell, at least in the early phase of 5G before device technology matures [1].

5G design targets to also bring the radio access point closer to the end device, thereby, shortening the last and most challenging segment of an end-to-end communication link and consequently reducing latency and increasing reliability. Many of the massive number of machine type communications can also benefit from the extended battery life resulting from shorter uplink distance.

Figure 2 shows an example of traffic distribution from both Smartphone-centric and router-centric LTE networks. This commercial network deployment example shows that the disproportionate traffic across cells may require to re-design network concepts in future evolution. It shows that in Smartphone-centric LTE network, 62% of users are camping on 38% of cells and generating 57% of traffic with an average user throughput being less than 6 Mbps; with 1/3 of cells generate ~ 50% of traffic to handle 60% of users. On the other hand, the router-centric LTE network has a traffic distribution where 66% of users are camping on 42% of cells and generating 59% of traffic with an average user throughput being less than 3 Mbps.

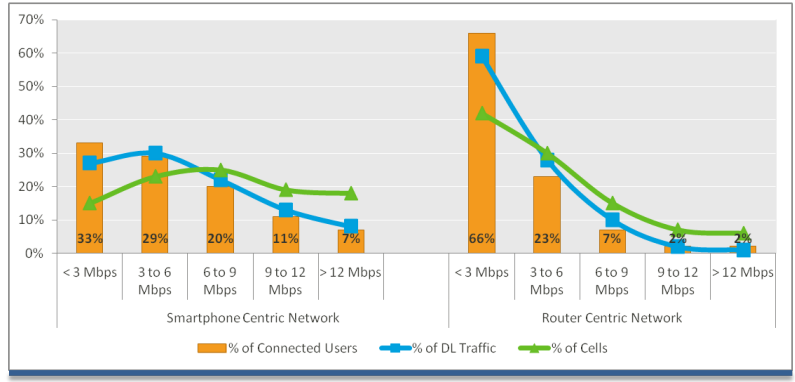

**FIGURE 2: LTE-A TRAFFIC & THROUGHPUT CORRELATION**

In the current LTE network deployment, the available use cases for connectivity leads to having concentrated traffic in a small area impact end-user experience. Moving towards small cells is important in order to have contiguous network with diversity of traffic and utilize a uniform network resources rather than deploying different networks to handle different types of traffic. Additionally, the Smartphone type traffic requires that the network to be dimensioned differently so that more cells are needed to ensure coverage everywhere while the cell resources are less utilized because of the bursty traffic nature of the Smartphone data connectivity. In this example, the overall downlink bit rate per Resource Block (RB = 180 KHz) is 0.22 Mbps only. The maximum theoretical bit rate per RB is 1.42 Mbps in LTE. This means that only 15.5% of a single RB throughput is utilized on downlink across cells in such network; leading to a waste of OFDM capacity.

To facilitate the deployment of a future mobile network that has small cell as its primary traffic bearing workhorse, the current radio access architecture needs to undergo some major revamps and new technologies need to be introduced.

Initially, the common layer could use LTE waveform (OFDM) but the dedicated layer uses the new 5G waveform. Therefore, 5G devices can be served by LTE for the locations where 5G service is not available. Over time, 5G coverage will be improved and 5G UE becomes popular. Therefore, the bandwidth of LTE-based network can be reduced or even replaced by 5G capabilities; which is the expected migration plan from a non-standalone deployment into a standalone one. This is discussed in next section.

## Massive Machine TYPE Communication

The Internet of Things (IoT) defines the way for intelligently connect devices and systems to leverage and exchange data between small devices and sensors in machines and objects. IoT concepts and working models have started to spread rapidly which is expected to provide a new dimension for services that improve the quality of consumer's life and the productivity of enterprises. The IoT effort started from the concept of Machine to Machine (M2M) solutions to use wireless networks to connect devices to each other and through the Internet, in order to deliver services that meet the needs of a wide range of industries.

Next to eMBB radio access, 5G will incorporate systems that enable massive machine-type communications (MTC). In 3GPP Release-13, NB-IoT was defined to operate within a 200 kHz bandwidth. In the work on 5G specifications, this is expected to be further optimized toward a high number of supported devices, low device cost, and ultra-low power consumption.

In order to support different types of deployments, NB-IOT targets three different modes of operations such as utilizing the spectrum currently being used by GERAN systems as a replacement of one or more GSM carriers (Stand-alone operation). The second mode utilizes the unused resource blocks within a LTE carrier's guard-band (Guard-band operation). The third mode utilizes resource blocks within a normal LTE carrier (In-band operation). The NB-IoT shall support the following main objectives:

- OFDMA on the downlink with 15 KHz subcarrier spacing
- SC-FDMA on the uplink with Single and Multi Tone of 3.75kHz and 15kHz subcarrier spacing
- A single synchronization signal design for the different modes of operation, including techniques to handle overlap with legacy LTE signals while reducing the power consumption and latencies
- Utilize the existing LTE procedures and protocols and relevant optimizations to support the selected physical layer and core network interfaces targeting signaling reduction for small data transmissions
- The supported deployment bands are: 1, 2, 3, 5, 8, 12, 13, 17, 18, 19, 20, 26, 28, and 66. Other bands not supported in Rel-13 are being studied to add for NB-IoT in REL-14

3GPP Release-14 introduces further enhancements to NB-IoT network and device capabilities in order to extent the solution to more use cases and applications [2]:

- Positioning Enhancements
    - Support of Observed Time Difference Of Arrival (OTDOA), or Uplink-Time Difference of Arrival (UTDOA) for better positioning accuracy, without adding significant device complexity or power consumption impact
- Multicast Support
    - Efficient software and firmware upgrade for massive devices with the introduction of enhancements to support narrowband operation
- Mobility Enhancements
    - Support connected mode mobility for service continuity for both user and control planes
- Lower Power Support
    - Lower Transmit power class (e.g. 14dBm) to support lower current consumption that are suitable for small form-factor batteries (e.g. for wearable devices)
- Higher Data Rate Support
    - Multiple HARQ process support and higher Transport Block Size (TBS) to increase downlink data rate from ~28kbps to ~112kbps

## 2. 5G NEW RADIO & AIR INTERFACE

The evolution of LTE in Release 14 is expected to offer a first step toward 5G by enabling wireless access for frequency bands below 6 GHz (sub-6). Hence, LTE Advanced Pro (LTE-A Pro) might be considered a special case of 5G in those frequency bands. For higher bands, a new radio-access technology (RAT) and inherent supporting and integration solutions will be introduced. Therefore, the 5G architecture will be an integration of Multi-RAT, supporting the simultaneous operation of multiple heterogeneous technologies.

It is therefore expected that 5G will initially work on three areas of improvements taking the LTE-A Pro design concepts into considerations:

- Radio Access that is providing service multiplexing for eMBB, mMTC and URLLC. This requirement is expected to provide scalable numerology and Flexible time-frequency grid. The target is to design a waveform that can be flexible for different sub-systems within the same carrier where the same spectrum resource to deploy new services;
- Radio Access that is designed to have lossless physical layer transmission reduce the pilot overhead and utilize connectionless transmission to reduce the control channel overhead, similar to that in NB-IOT. This requirement is expected to provide a reduction in overhead and achieve high efficiency in 5G networks. In the current LTE deployment, the overhead of control and pilot channel in 20 MHz bandwidth deployment can reach up to 28%. Therefore, 5G may aim at reducing the transmission overhead of the control/pilot channels, restricting the necessary overhead in the narrowband for initial access and configure dynamically a wider bandwidth operation for eMBB;
- Radio Access that is capable of low latency transmission. This can be achieved by having shorter OFDM symbol length, shorter Transmission Time Interval (TTI), contention-based uplink, and a modified carrier spacing in order to meet latency requirements of < 10 ms for eMBB services.

Therefore, 5G is a portfolio of access and connectivity solutions addressing the demands and requirements of mobile communication for a wide range of services and applications. Current and future mobile networks have to overcome several challenges:

- How to manage highly diverse deployment strategies and topologies;
- How to maintain a consistent user experience across all network layers and locations in densely interference environments;
- How to reduce cost per bit, increase the capacity and at the same time maximize the return on investment.

In order to accelerate the next generation mobile technology that is capable of meeting these challenges, a coexistence with LTE-A Pro (eLTE in its Release-15 version) network is expected to be a key enabler to 5G. There is strong industry interest in completing the non-standalone (NSA) version of the 5G new radio specifications on the basis of the legacy LTE architecture (EPS) before March 2018. Several options are under discussions including the ones shown in figure 3. The deployment strategy (among 11 possible strategies) suggests that the 5G design will consist of Next Generation Radio (or New Radio – NR) at the access side and Next Generation Core (NGCN) at the core network side. These two entities can be deployed as standalone or combined with LTE radio and core network. 3GPP is taking the direction of independent radio and core migrations whereby NR and NGCN do not necessarily come together. The key aspect of NGCN will be slicing. Therefore, some operators may want slicing for accommodating new businesses without introducing a new radio (i.e. use the existing LTE radio but with NGCN).

The NSA architecture will have a radio access that is LTE assisted (LTE as an anchor layer) with 5G radio in a dual connectivity mode, while the core network will remain on the top of the legacy Evolved Packet Core (EPC) as in LTE [3].

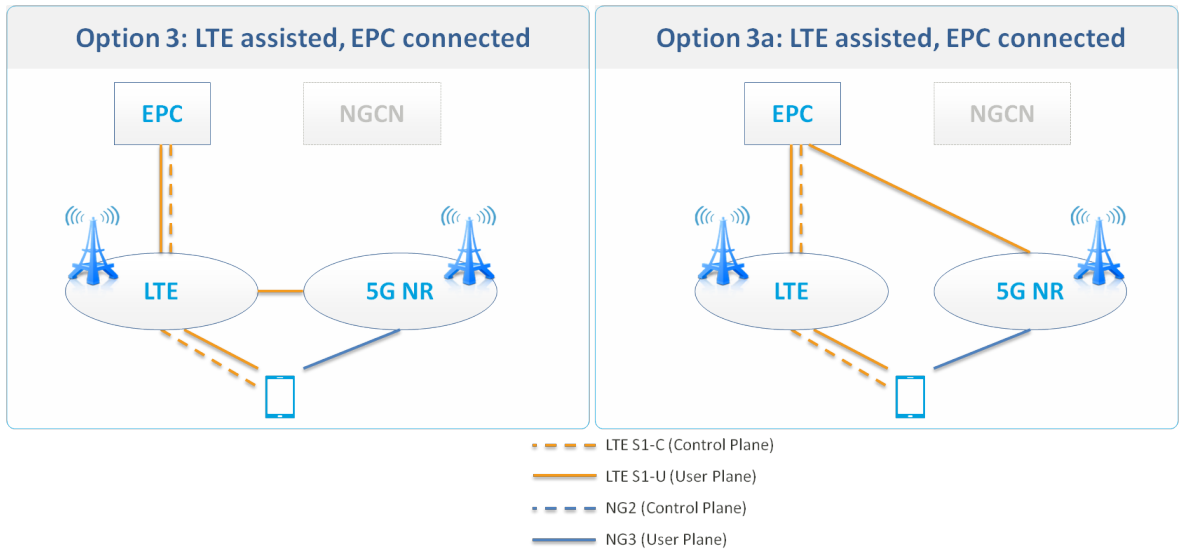

FIGURE 3: 5G MIGRATION SCENARIOS IN 3GPP – NON-STANDALONE (NSA) OPTIONS

On the other hand, it is also affirmed that there is another strong industry interest in completing the Standalone (SA) option 2 and option 4/4a/5/7/7a by the agreed deadline of June 2018. Figure 4 shows the possible deployment options under further study now. However, both cases will initially deal with eMBB and URLLC related use-cases [3].

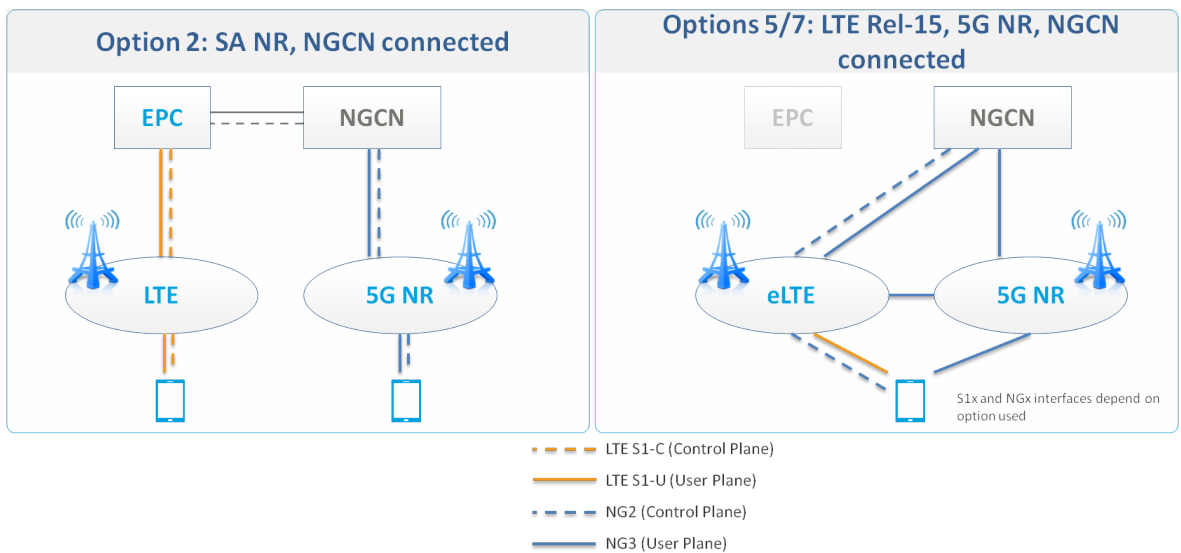

FIGURE 4: 5G MIGRATION SCENARIOS IN 3GPP – STANDALONE (SA) OPTIONS

## What is Next for LTE-A Pro Evolution?

From now until 5G realizes real deployment tractions in 2020, much of the available mobile network coverage will continue to be provided by LTE. Because the 5G will most likely coexist with LTE and other technologies such as Wi-Fi access, it becomes important that operators with deployed 4G networks have the opportunity to manage the existing network efficiently and provide a good underlying access layer into 5G, especially for NSA type 5G deployment.

The evolution of LTE to LTE-A introduced three main categories: carrier aggregation above 20 MHz bandwidth, higher order modulation beyond 64QAM, and higher order MIMO (Multiple Input Multiple Output) beyond 2x2. These features can be deployed in order to improve the peak data rates and spectral efficiency.

The theoretical peak data rates increased from 300 Mbps in the downlink and 75 Mbps in uplink (Release 8) to 3 Gbps in the downlink and 1.5 Gbps in the uplink (Release 12). The most important feature LTE-A introduced to meet those requirements was carrier aggregation (CA) to enable peak data rates above 150 Mbps on downlink and above 50 Mbps on uplink. The current common deployment uses up to three component carriers (CCs) in downlink and two CCs in uplink with up to 450 Mbps and 100 Mbps. Furthermore, 3GPP specifies MIMO extensions to 4x4 in the downlink and also add higher order modulation with 256QAM on downlink and 64QAM on uplink.

LTE-Advanced Pro maximum downlink data rates are expected to exceed 600 Mbps when combining these features together as in Category 15 and 16 deployments shown in various scenarios in figure 5.

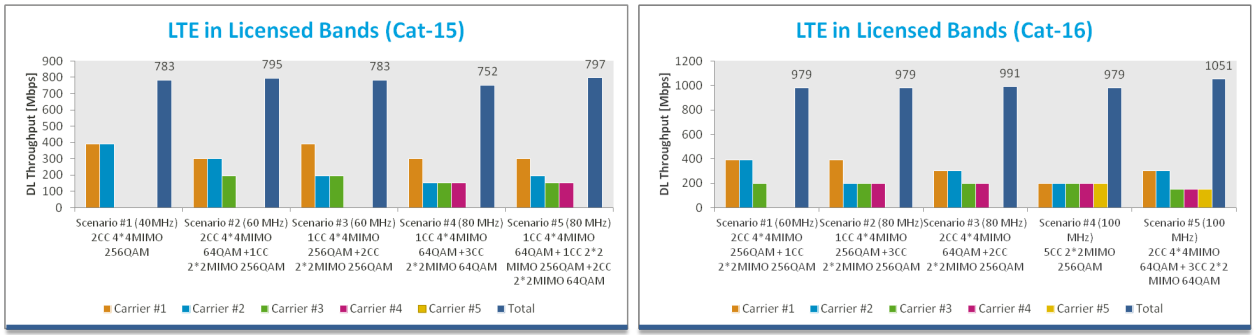

FIGURE 5: LTE-ADVANCED PRO PEAK THROUGHPUT DEPLOYMENT SCNEARIOS

Packet latency is another performance metric used by mobile network operators and end-users to measure end-to-end quality of service. There are many existing applications that would benefit from reduced latency by improving perceived quality of experience, including real-time applications like Voice over LTE (VoLTE), video telephony, and gaming. Furthermore, the number of delay-critical applications will increase: we will see remote control and autonomous driving of vehicles, augmented reality applications, and specific machine communications requiring low latency as well as highly reliable communications. 3GPP has specified work items in order to improve latencies in LTE network with concepts that are expected to be carried over to 5G developments at a later stage.

All these improvements in LTE-A deployments will reflect positively into the evolution to 5G. It is therefore good to summarize the current gaps in 4G LTE network and how 5G will bridge them over the next few years.

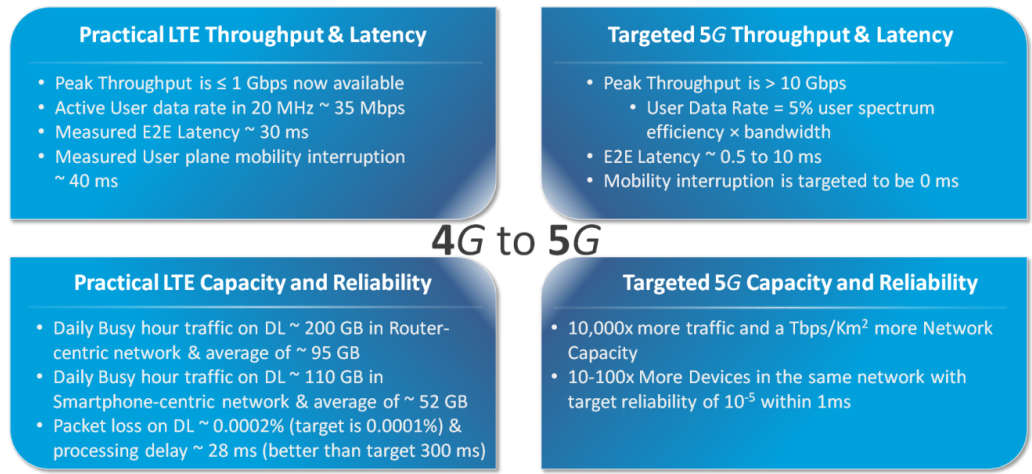

**FIGURE 6: 5G AND LTE GAP ANALYSIS**

## 3. 5G SPECTRUM VIEW

One of the significant design concept changes coming into 5G is enabling cellular transmission with all types of spectrum and bands to support a wide range of new services with different deployment requirements. In order to meet the demand of the increasing traffic capacity, enable the transmission bandwidths needed to support very high data rates at one end, and a diversity of use cases at the other end, the 5G design will extend the range of frequencies used for cellular bands. This includes utilizing new and existing spectrum below 6GHz (sub-6), as well as defining new spectrum for cellular use in higher frequency bands (above 6GHz).

The standardization and regulation bodies worldwide are defining 5G roadmaps with input from 4G Americas including USA, 5G Forum Korea, 5GMF Japan, 5G-PPP Europe, and the IMT-2020 5G Promotion Group China. The typical alignment mostly takes place between ITU-R (the International Telecommunication Union, Radiocommunication sector) and 3GPP. In early 2012, ITU-R initiated a program to develop International Mobile Telecommunication (IMT) system for 2020 and beyond (IMT-2020), thereby officially kicked off the global race toward a yet to be defined 5G mobile network. The vision of this next-generation system began taking shape with ITU-R WP5D by proposing a work plan on spectrum and technology timelines [4].

The key ITU-R IMT-2020 roadmap is shown in figure 7. ITU-R will open the evaluation criteria as of October 2017, open the window of proposal submissions as of June 2019, and finally set the specification details as of October 2020. In the meantime, 3GPP set up its roadmap to address 5G into two phases, with the final phase in 2019 having its specifications ready for submission to ITU-R as part of IMT-2020 in February 2020. 3GPP started work on 5G in September 2015. The 3GPP specifications for 5G will come in two phases:

- 5G Phase 1 to be completed towards the end of 2018 (3GPP Release-15) and set up the priorities for:
    o Spectrum and waveforms up to 40GHz for both eMBB and low latency use cases
    o Radio Migration to/from LTE as discussed in previous section with Non-standalone being the feasible operation, network slicing, mobility session management, basic core network policies and security, and IMS (voice/SMS)
- 5G Phase 2 to be completed towards the beginning of 2020 (3GPP Release-16) and set up the priorities for:
    o Spectrum and waveforms above 40GHz and adding mMTC, URLLC for Vehicle to Vehicle/Everything (V2V and V2X) use cases
    o Shared and unlicensed spectrum including interworking with other cellular systems
    o Additional uses cases and services including proximity services, multimedia broadcast services, public warning/emergency alert, Satellite communication, etc..

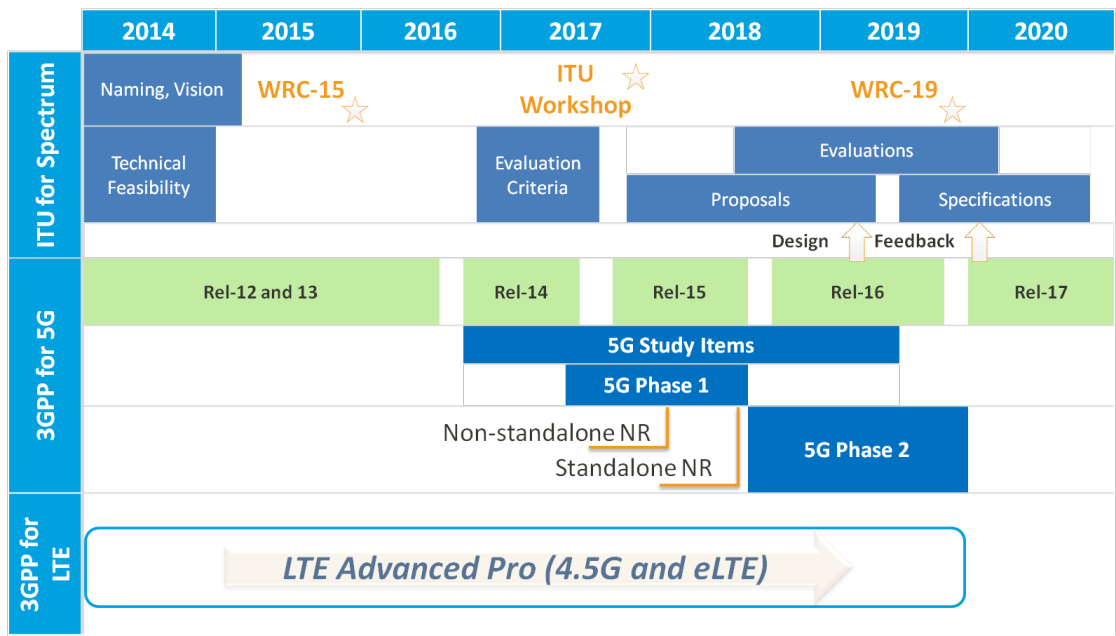

FIGURE 7: TIMELINE FOR IMT-2020 (5G) DEVELOPMENT

From now until the first phase of 3GPP is completed, the industry needs to address the available spectrum and technical capabilities to address the initial stage of deployment. Therefore, the ITU World Radio communication Conference 2015 (WRC-15), addressed extra level of harmonized spectrum for different industries including mobile and wireless communications. One of the key achievements in this context was the allocation of an additional IMT spectrum within 470 MHz to 6 GHz. For example WRC-15 defined the largest contiguous range of 200 MHz between 3400 and 3600 MHz, known as C-Band. Then in WRC-19, it is expected that it will deal with the range of bands above 6 GHz for IMT-2020 within the 24.25-86 GHz, as shown in figure 8.

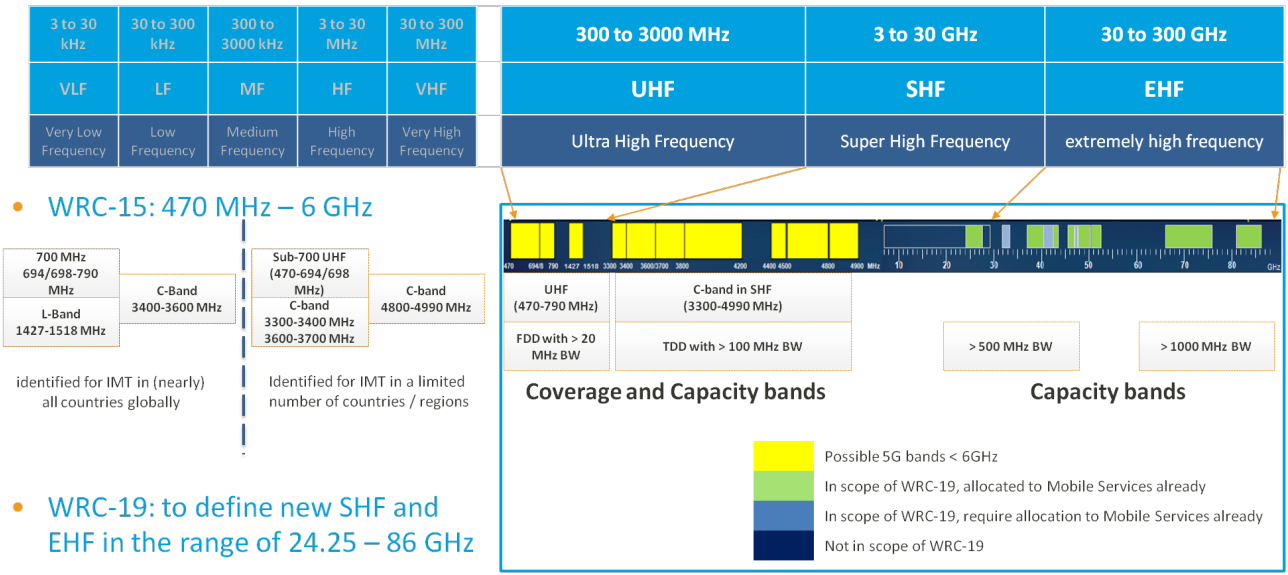

**FIGURE 8: WRC SPECTRUM DEFINITIONS**

## 4. UAE 5G INNOVATION GATE (U5GIG) INITIATIVE

The U5GIG has been envisioned to be a consortium of technical and academic organizations in the UAE as well as global telecom vendors to plan and use their expertise to define and develop a global 5G network that will radically change lives across the United Arab Emirates (UAE). In addition, U5GIG allows universities and technical organizations across the UAE to work together and participate in the development of the 5G ecosystem, and for academia and industry to test applications and technologies in a real-world setting. Moreover, U5GIG will deliver innovative communications solutions in order to generate social and economic value. As a result, U5GIG puts UAE at forefront of mobile innovation and at the heart of networking development.

du is taking the lead to build a UAE 5G Innovation Lab to prototype, test and validate early 5G and IoT equipment and services. The aim of this initiative is to bridge the gap between telecom industry and academia in UAE by establishing and maintaining close, productive collaborations with academic institutions, industry and the community. The ultimate goal is to adopt a collaborative approach to the development of the 5G ecosystem and assess 5G solutions via real-life smart use cases and applications.

In order to achieve this, du, as well as other consortium members, including MediaTek, will jointly supervise 5G research programs with the major UAE universities and based on carefully selected practical 5G research topics. Accordingly, the UAE will contribute to and participate in the standardizations and development of 5G in forums such as 3GPP, ITU, and GSMA. We plan to work closely with suppliers and SMEs and eventually train future UAE academic and industry leaders to bring UAE's voice to the technology development debate.

U5GIG will host vast number of IoT use cases and IoT platforms to have an open forum for application development in UAE. We are proud to build the 1st of its kind open standard 5G and IoT lab in UAE for applications and use cases development. UAE enterprises and universities will have access to this great and unique innovation center to develop customized and innovative IoT applications. We will have three streams in the 5G innovation lab: massive MIMO stream for sub-6 GHz, millimeter wave (mmWave) stream with 3D beamforming, and IoT stream. The Massive MIMO and mmWave streams will realize the extreme broadband experience of the 5G while the IoT stream will build and develop IoT use cases and application on top of existing and new technologies.

## 5. DU 5G AND IOT ROADMAP

du has rich wireless and fixed broadband portfolios. Figure 9 summarizes the portfolio of du's broadband technologies. 5G is expected to be a unified network that can cater for different use cases. The 5G access network will be designed in a flexible manner to accommodate different application requirement such as extreme mobile broadband, Multi-Giga Fixed Wireless access, massive MTC and URLLC IoT applications and use cases. du offers comprehensive broadband connectivity that covers different segments and different applications. At du we always search for new technologies and we are leading the region in terms of technology innovation and introduction of new technologies.

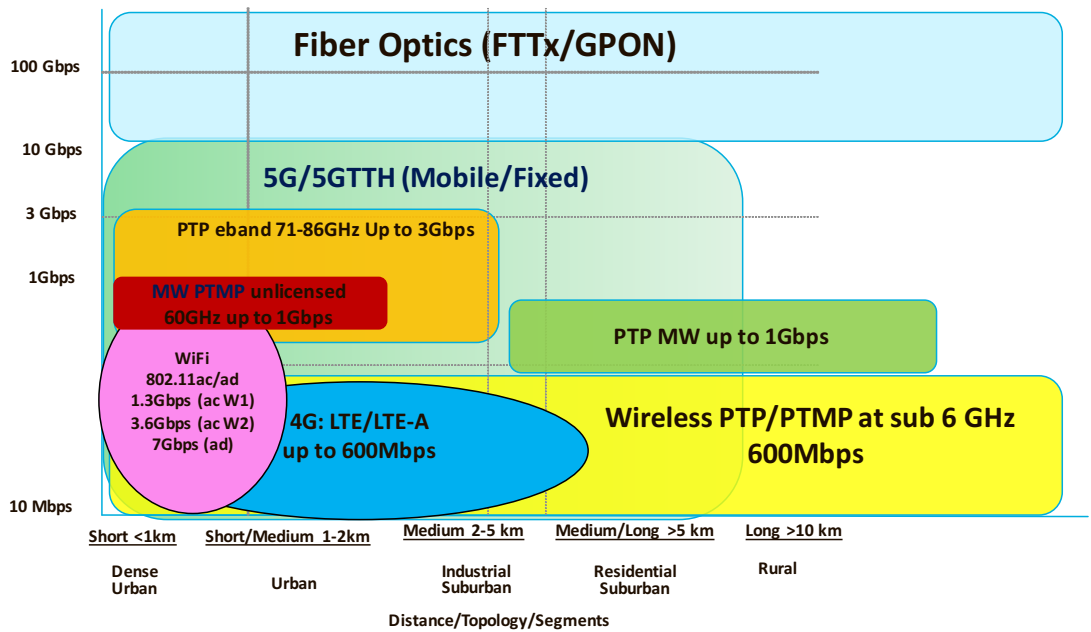

**FIGURE 9: PORTFOLIO OF DU'S BROADBAND TECHNOLOGIES**

Having said that, du is leading the mobile broadband evolution in the Middle East. Du's major milestones in the journey of mobile broadband are summarized as follows:

- 5G prototype based on mmWave with peak TP of 10 Gbps (highest throughput in the region) in GITEX 2015;
- LTE-A with peak throughput of 900Mbps with three component carrier (3xCC) and MIMO 4x4 in GITEX 2014;
- LTE-A with Carrier Aggregation, and another one with LTE MIMO 4x4 demonstrated with peak downlink of 300Mbps in GITEX 2013, 2012, respectively;

- LTE 150 Mbps with CAT 4 devices and demonstrated it at GITEX 2011 at the same time the LTE network has been commercially deployed;
- DC-HSPA+ with peak TP of 42Mbps at GITEX 2010.

du has successfully tested and demonstrated the first 'true' IoT network in the Middle East using unlicensed LPWA network in 2015. Several use cases were demonstrated such as smart parking, smart environment and smart tracking. In addition, we are currently testing 3GPP NB-IoT technology for future IoT use cases. du have successfully built several IoT use cases on existing technologies (e.g. 2G, 3G, LTE, Wi-Fi, Zigbee, PLC, and LPWAN) including smart parking, smart meter, smart waste management, smart environment, smart building, smart lighting, and smart fleet management. From the experience gained in deploying such platforms, the main challenge in IoT implementation is building an end-to-end use case and being ready with the required infrastructure to provide bottom-up approach for use cases including radio connectivity, sensors, applications, IoT platform, analytics, big data mining and machine learning. This is a fundamental part of our integrated digital strategy to go beyond connectivity with the strategy set as follows [5]:

- Offers multiple and hybrid technologies according to the use case requirements including: throughput, coverage, power, latency, cost, and spectrum;
- The existing networks (2G, 3G, LTE, Wi-Fi) will meet the applications that need long range and high data rates, expected to constitute 10% of the IoT market volume,
- Technologies such as ZigBee, RF Mesh (802.15.4), PLC, Wi-Fi will be used for short range applications such as smart meter, smart home, smart parking , expected to constitute 30% of the IoT market volume;
- Introduce Low Power Wide Area Network (LPWAN) based on 3GPP's NB-IoT for nationwide use cases. This will be the main stream for critical IoT applications while the LPWA unlicensed network can be used for non-critical application and for on specific use cases, expected to constitute 60% of the IoT market volume.

## 6. CONCLUSION

3GPP has started the 5G discussion in RAN 5G Workshop September 2015, and has reached a substantial progress since then, consolidated an accelerated timeline for early NR deployments and is on-course to start normative work on the overall architecture in December 2016 and on the new radio technology in March 2017. As defined to be a unified system to cover a wide range of use cases, 5G is expected to cover a diversity deployment that is scalable and adaptable for different spectrum types; low and high bands. It is generally agreed by the industry that 5G standardization will take two main releases, spanning over Release-15 and Release-16. The first phase will prioritize spectrum and waveforms up to 40GHz for both eMBB and low latency use cases, while the second phase will continue with waveforms above 40GHz and adding mMTC and URLLC use cases. At the same time, ITU-R extended the sub-6 GHz bandwidth for IMT at a nearly global scale in WRC-15 covering new range of bands like the C-Band, while considering the 24 to 33GHz spectrum for mmWave technologies as the range considered for the potential global harmonized band. The envisioned market space for 5G technology is targeting a design that has the capability to unify the system needed by various use cases. In order to enable services for a wide range of users and industries with new requirements, the capabilities of 5G must extend beyond those of previous wireless access generations. These capabilities will include enhanced mobile broadband connectivity, massive system capacity for machine type communications, very high data rates everywhere, very low latency, ultra-high reliability and availability, and low device cost. 4G has transformed the internet making it mobile and with it enabling countless services to flourish. 5G will leverage this further and make internet truly ubiquitous whilst enabling new disruptive use cases even beyond what is feasible today. Therefore, in order to accelerate the next generation mobile technology that is capable of meeting these requirements, a coexistence with LTE-A Pro network (eLTE in its Release-15 version) is expected to be a key enabler to 5G deployments.

MediaTek and du are dedicated to the success of 5G. With a technology leadership spanning through multimedia, mobile communications, connectivity and computing technologies – all of which set to play an essential role in bringing 5G to life by 2020, MediaTek and du are well positioned to be a pioneer of the brand new world enabled by 5G.

## REFERENCES AND FURTHER READING

## ABOUT MEDIATEK

The World-class team of experts at MediaTek has been actively involved in exploring, defining and validating technology for the fifth-generation mobile communications system, whilst also engaged in related local and international collaboration efforts and fully committed to its timely standardization in 3GPP. In order to prevent technology fragmentation, to guarantee competition and compatibility, to channel investments and enable economies of scale for operators and users alike, global standards are needed. MediaTek is therefore dedicated and committed to bring 5G to reality by 2020.

## ABOUT DU

du has rich wireless and fixed broadband portfolios. du offers comprehensive broadband connectivity that covers different segments and different applications. At du we always search for new technologies and we are leading the region in terms of technology innovation and introduction of new technologies. du is taking the lead to contribute to the 5G ecosystem by building a UAE 5G Innovation Gate (U5GIG) lab to prototype, test and validate early 5G and IoT equipment and services. The aim of this initiative is to bridge the gap between telecom industry and academia in UAE by establishing and maintaining close, productive collaborations with academic institutions, industry and the community. The ultimate goal is to adopt a collaborative approach to the development of the 5G ecosystem and assess 5G solutions via real-life smart use cases and applications.